\documentclass{article}
\usepackage{amssymb}

\input{tcilatex}

\begin{document}

\title{The Limits of Quantum Mechanics in the Light of the Genuine
Fortuitousness Principle}
\author{Jerzy Czyz \\
Montana State University - Billings, Department of Mathematics}

\begin{abstract}
To test the limits of quantum mechanics, a proposal for an experiment on
protons is suggested. \ The spin component of proton is measured very
rapidly in sequential measurements. \ The reason for this experiment is
derived from the genuine fortuitousness principle, (A. Bohr, B. R.
Mottelson, and O. Ulfbeck, (2004)).

\ 
\end{abstract}

\begin{center}
{\Large \bigskip The Limits of Quantum Mechanics in the Light of the Genuine
Fortuitousness Principle}

{\small Jerzy Czyz}

Montana State University - Billings, Department Of Mathematics
\end{center}

{\small \ }

\bigskip

This paper is concerned with the limits of quantum mechanics (QM). \ In
point 4, we put forward a proposal for a thought experiment on protons. \ It
is our opinion that an experiment of the described type must yield results
which deviate from the predictions of QM. \ Additionally, the first three
points take a sober look at some basic concepts of QM to expose the idea and
motivation behind our proposal in a pedagogical way. \ The following
methodological principle is accepted: What belongs to QM (as a physical
theory) is only what is needed to derive it. \ We hope that the reader will
find this particular application of the Occam's razor quite useful.

It is the basic assumption of QM that%
\begin{equation}
tr(DP_{A}(E))
\end{equation}%
is the probability that the value of the physical quantity represented by a
self-adjoint operator $A$ (acting on a suitable Hilbert space), when the
state of the system is the density operator $D$, lies in the Borel set $%
E\subset R$. \ The function $E\rightarrow P_{A}(E)$ is the spectral
projection-valued measure of $A$. \ If $A$ and $D$ are fixed, the function%
\begin{equation}
E\rightarrow tr(DP_{A}(E))
\end{equation}%
has the properties of a probability measure on $R$. \ Many vexing questions
and paradoxes arise, when ontological assumptions on the nature of states,
physical quantities, quantum systems etc. are incorporated into the
formalism of QM.

1. \ \textit{Critique of operational approach to states and physical
magnitudes.} \ The reader can find a concise exposition of the operational
approach to states and observables in Araki (1999). We wish to point out
some omitted assumptions and other imperfections there. \ Let us start with
a short review of the operational approach. \ QM presupposes the division of
the world into three parts:

- a physical system (entity) entering into a measurement process

- a measuring apparatus which must be coupled with the physical entity to
make a measurement

- the rest, besides the above, called the environment and observer.

What is taken to be a physical system is beyond the reach of full
mathematical formalization. \ It is presumed that a part of reality can be
cut out in such a way that physical theory applies to it. \ Many physical
systems have hierarchical structure, and permit sensible divisions into two
or more subsystems, atoms for example. \ But there are systems like an
electron or a neutrino for which we are not able to devise a sensible
decomposition into subsystems. \ In conclusion, a specification of a
physical system is to a large degree a matter of convention, although not
completely arbitrary. \ We will return to this issue one more time later.

Physical quantities of QM cannot be characterized by a single numerical
value, in general, but by a whole set of possible values the physical
quantity can take. \ The validity of the above follows from realization that
the result of a single one-time measurement on the system prepared by the
most discerning preparation procedure empirically available, and ultimately
controlled arrangement between measuring instrument and measured system
cannot be predicted with certainty. Acausal traits of QM enter the scene
here. If this is the case, then one is left with the following empirical
solution: repeat the measurement many times on many individual systems
believed to be prepared in the same way as the physical system in the first
measurement. \ The reader is asked to keep in mind that the repetition (in
different times or places simultaneously) necessary involves assumptions
about space-time. \ The operational approach is not possible without these
assumptions. \ The finite sequence $(l_{1},l_{2},...,l_{N})$ gives the
record of the measurement of physical quantity. \ Assuming for simplicity
that the measurement takes only discrete values $q_{1},q_{2},...$ one can
compute the relative frequency $f_{i}$ of the occurrence of the value $q_{i}$
in the record:%
\begin{equation}
f_{i}=\frac{n_{q_{i}}}{N}
\end{equation}%
where $n_{q_{i}}$ is the number of occurrences of $q_{i}$ in the record. \
We bore the reader with these trivialities to stress the point that for an
elusive quantum system these are the only empirical data we have. \ Having
in mind these elusive systems, the first step towards theory is to talk
about records and frequencies in the probabilistic way. \ It is an error to
say that if%
\begin{equation}
\lim_{N\rightarrow \infty }\frac{n_{E}}{N}
\end{equation}%
exists, then one looks for a probability measure such that%
\begin{equation}
p_{\alpha }^{Q}(E)=\lim_{N\rightarrow \infty }\frac{n_{E}}{N}
\end{equation}%
for every set $E\in B(R)$ , where $B(R)$ is $\sigma $-algebra of Borel sets
on real line, and LHS is the probability to find the value of the physical
quantity $Q$ in the set $E\in B(R)$ , where $\alpha $ is the `state' of the
system.

To begin with, the RHS of (5) depends on infinitely many results in the
future. \ Secondly, the frequentist definitions of probability have well
known serious mathematical defects, and cannot serve as a basis of a
probability theory. \ The reader can consult C. Caves and R. Schack (2004)
to see the total bankruptcy of frequentist approach in QM. \ We rewrite (5)
in the following form%
\begin{equation}
p_{\alpha }^{Q}(E)\longleftarrow \frac{n_{E}}{N}
\end{equation}%
where $\longleftarrow $ indicates that RHS serves as the method of
estimation of the theoretical distribution $p_{\alpha }^{Q}(E)$ based on the
histogram obtained from the record of the measurement. \ In this way we
reach the first preliminary conclusion: in orthodox QM, the mathematical
probability theory (i.e., measure theory) on the lattice of subspaces of a
Hilbert space is primary. \ Records and frequencies (3) serve to verify
(falsify) this theory. \ The above conclusion imposes a severe limitations
on all discussions about the nature of probabilities in QM. \ The
probabilities belong to mathematics. \ The theoretical prediction (2) can be
verified (falsified) by histograms obtained from \ (3). \ This relation has
a similar character to the relation between points and lines in Euclidean
geometry, and material points and lines on a piece of paper, or in the sand.
\ To a mathematical line of geometry one can correlate material line on the
paper or in the sand, and continue in this way analysis of drawn polygonal
figure. \ All discussions about the nature of the line belongs to geometry.
\ If one thinks that the Euclidean line is a limit case of real material
line, then one commits a serious error. \ 

To make formula (6) more precise, discussion of states and physical
quantities is needed. \ In idealized scheme of Araki (1999), two individual
physical entities (systems) $\mathbf{s}_{1},\mathbf{s}_{2}$ are equivalent ($%
\mathbf{s}_{1}\sim \mathbf{s}_{2})$ if%
\begin{equation}
p_{\mathbf{s}_{1}}^{\mathbf{A}}(E)=p_{\mathbf{s}_{2}}^{\mathbf{A}}(E)
\end{equation}%
for all $E\in B(R)$ , and all $\mathbf{A}$ , providing that $\mathbf{A}$ is
real measuring procedure (apparatus etc.) applicable to both individual
systems. \ This relation $\sim $ is an equivalence relation. \ If $\mathbf{s}%
_{1},\mathbf{s}_{2}$ belong to the same equivalence class, then they cannot
be distinguished by any physical quantity $\mathbf{A}$ . \ The crucial point
here is that mathematical entities called states are attached to these
equivalence classes and not to individual systems.

We must admit the possibility that two different instruments $\mathbf{A}_{1}$%
and $\mathbf{A}_{2}$ can be coupled with every system described by the state 
$s$ to measure the same physical quantity. If for all states $s$ and all $%
E\in B(R)$%
\begin{equation}
p_{s}^{\mathbf{A}_{1}}(E)=p_{s}^{\mathbf{A}_{2}}(E)
\end{equation}%
holds, then $\mathbf{A}_{1}$and $\mathbf{A}_{2}$ measure the same physical
quantity. \ This defines an equivalence class of instrument $A$ .This
concludes the operational definition of $p_{s}^{A}(E)$ . \ The state
characterized above by (7) and (8) can be called, after Bub (1988), a
statistical state, and it is an additional assumption that it is hereditary,
possessed by each individual system $\mathbf{s}$ described by the state $s$
. \ However, we do not want to discuss this any further. \ Instead, we want
to point to a fundamental assumption which underpins the whole discussion. \
The elusive physical system is not given directly. \ What we have at our
disposal are instruments which measure physical quantities. \ The phrase
`the state of individual physical system is $s$ ' presupposes the fusion
(stitching) of all relevant physical quantities into one theoretical
structure which pertains or can be attributed to an individual system. \
Without this fusion principle there are no entities like quantum systems. \
Again, it is additional assumption to say that after performing this fusion,
one arrives in entities with possessive attributes. \ Imagine a man blind
from birth. \ He has the natural view of the world based on impressions from
other senses. \ When he regains his sight after successful surgery, he does
not see red apples and green leaves, but chaos of colors. \ After some time
he will organize his visual experience, and fuse it with his previous
non-visual natural knowledge of the world - beforehand assumptions about
structure of the world. \ In this \ case the fusion is such that it does not
destroy his believe in things. \ Hence, it is a mistake to assume in advance
that after applying fusion to relevant physical quantities, one will arrive
at theoretical structure put beforehand. \ In conclusion, the operational
definition of a state is not possible because of three main reasons: some
knowledge about space-time must be assumed, mathematical probabilities are
primary in interpretation of records, and the definition of a quantum system
is based on an application of the fusion principle.

2. \textit{Second look at quantum systems. \ }We stated earlier that
formalization of the concept of \ a quantum system is beyond the reach of
mathematical formalism. \ Suppose that an experimentator works with \ a
single electron. \ What is the theoretical correlate of an electron? \ The
answer to this question has been given by Wigner (1939), (1959). \ Omitting
details, the response is this:

-Quantum systems are given by irreducible unitary representations of the
group $G$ of physical theory. \ Additionally, the group $G$ must satisfy
certain constraints imposed by physics.

-Group $G$ is double covering of the Poincare group $P$ of those
transformations of the special relativity which can be continuously derived
from identity or $G$ is enlarged to account for quantum numbers. \ For
example, group $SU(3)$ is used for such enlargement.

This characterization of quantum system is connected with symmetries of the
whole Minkowski space-time but not with a particular quantum system. \
Additionally, the idea that \ unitary representation is intrinsic to a
quantum system is absurd. \ To achieve the complete specifications of his
quantum system, the operationalist must rely on localization of events in
the space-time. \ One can use a positive-operator-valued measure (POVM) on
the Poincare group to determine the probability that a measurement of
coordinates of the event gives a result belonging to a given set in
space-time, S. Mazzucchi (2000), M. Toller (1998). \ However, such a
specification of quantum system is possible only if relevant observables in
the formalism of POVM are fused together to relate to the individual quantum
system.

If the possibility of states and observables to quantum entities is denied,
operational view of QM is considered to be untenable, and the concept of
quantum system is put in doubt, then, at first glance, there is nothing
left. \ However, this is not the case. \ The missing basic principle has
been proposed by A. Bohr, B. R. Mottelson, and O. Ulfbeck (2004). \ In the
next point we give a short description of this principle because the idea of
the proposed experiment stems from it.

3. \textit{The Principle of Genuine Fortuitousness. \ }QM presupposes the
existence of instruments and their records. \ The relation between the
records and instruments is the relation of precedence (in the sense that a
mother's birth absolutely precedes the birth of her child) more fundamental
then time durations, and independent of any observer. Therefore, QM
presupposes existence of some knowledge to deal with experimental
situations. \ The requirement that this knowledge has a form of classical
physics is additional, and not needed further. \ This knowledge is such that
it is independent, and does not involve QM. \ We will call it NonQMTh. \ For
instruments with the highest resolution, we accept (Bohr, Mottelson, Ulfbeck
\ (2004)) the principle of the click.

A click without a precursor, totally acausal, not permitting further
analysis, unfathomable in its nature occurs in the counter of apparatus. \
Further work of apparatus leads to an entry in the record of the
measurement. \ An entirely acausal click is recognized as a macroscopic
event, which can be characterized as a discontinuity in space-time. \ The
concept of impinging particle or other entity `acting' on the counter is
denied. \ The formalism of QM is set in space-time pocked by clicks. \
Experimentators organize the clicks into sets. \ Such sets yield records. \
In orthodox QM the theoretical statistics of the set of clicks comes from
the triple 
\begin{equation}
(\rho ,A,tr(\rho A))
\end{equation}%
where $\rho $ is a density operator on suitable Hilbert space $H$ ( a
linear, bounded, self-adjoint, positive, trace-class operator on $H$), and $%
A $ is self-adjoint operator on $H$ . \ The physical sense of the triple is
given by%
\begin{equation}
tr(\rho P_{A}(E))\longleftarrow \frac{n_{E}}{N}
\end{equation}%
for any finite record $(l_{1},l_{2},...,l_{N}),$ where $P_{A}(E)$ is
projection operator corresponding to $E\in B(R)$ . If the click principle is
accepted, then the whole formula (10) makes sense: it deals with stochastic
character of sets of clicks. \ There are no electrons or other fundamental
quantum systems as objects or entities producing (causing) events. \ They
are symbolic structures assembled in the framework of the theory. \ These
symbolic structures are characterized by quantum numbers, but quantum
numbers do not belong to particles or things.

At a very fundamental level (the highest possible resolution of our
instruments) there is only theory of no things. \ When the resolution of
instruments decrease, then symbolic structures of theory come closer to the
concept of a thing, (Bohr, Mottelson, Ulfbeck \ (2004)).

Before we propose our experiment a few words are in order on mutual relation
between QM and NonQMTh. \ The principle of genuine fortuitousness \ assumes
the very basic type of the instrument and physical quantity connected with
it. \ NonQMTh is based on conceptual structures quite different from those
of QM. \ The properties of an apparatus (complex system of large number of
parts and high degree of organization) are described in a rather small
number of terms with no relevance to QM. \ Therefore, the theoretical
description of the measurement process should revolve about definitions and
construction of global observables (macroobservables) which can be later
`identified' with macroobservables of NonQMTh. \ It is unavoidable that such
a theory must be based not only on QM (orthodox or algebraic) but also on
structures that account for characteristic features of the instrument. \ The
reader can consult Sewell (2002), (2005) for details.

4. \ \textit{Limits of QM, and a thought experiment to test it. \ }We do not
consider QM to be the final theory, and because of that we will attempt to
find an experiment which will falsify QM. \ The previous discussion is at
odds with all doctrines of hidden variables. \ Hence, one must look in a
different direction. \ If essentially acausal character of QM is considered
to be primary, then the main possibilities are:

-There are sets of clicks not discovered yet, which do not obey the laws of
QM (orthodox or algebraic), especially formula (10) is violated. \ They can
be discovered using instruments of current quality and resolution.

-Spatiotemporal description by Minkowski space-time fails, for example the
Lorentz invariance is violated. \ The quantum systems support irreducible
representations of the Poincare group. \ Therefore, it is reasonable to
expect that an insufficiency of Minkowski space-time will have far reaching
consequences for QM.

-When the quality and resolution of our instruments are pushed forward,
statistical prediction of QM do not hold.

We want to consider the last possibility. \ We begin with the simple
observation, that the natural view of the world depends on the speed with
which \ a living organism receives impressions. \ Living organisms are
capable of receiving only finite number of impressions per minute (second).
\ In consequence, phenomena of such duration that they fall between two
consecutive impressions are not perceived, and the natural view of the world
is not only relative but false.

Similarly, the statistical regularities of clicks described by QM depend on
speed with which an apparatus is able to receive clicks. \ It is feasible
that all quantum measurements carried out so far missed some collections of
clicks because of their limitations. \ This calls for an experiment with
very rapid successive measurements. \ We propose an experiment, where an
individual system is subjected to a series of sequential measurements.

Specifically, the spin of a proton will be measured sequentially by two
devices A and B. \ Two measurements are separated by the time interval $%
\Delta T$ . \ Each device measures the spin component of a proton in its own
direction. \ For simplicity, let us assume that the first device measures
the spin component along $z$ axis, and the next device measures the spin
component along axis $z^{^{\shortmid }}$ with the angle $\theta $ between
them. \ Both are ideal von Neumann measurements. \ The initial state of a
proton is $\psi $. \ If the device A gives value $a_{m}$, then $P_{a_{m}}$
is appropriate state for the device B. \ Hence, we have the known formula
for the probability that the measurement by B will give result $b_{n}$ :%
\begin{equation}
prob(B=b_{n}\mid A=a_{m})=tr(P_{a_{m}}P_{b_{n}})=\mid (b_{n},a_{m})\mid ^{2}
\end{equation}%
This probability does not depend on $\psi $ and $\Delta T$ .

\textit{Hypothesis}: If the time separation $\Delta T$ between two completed
consecutive measurements is close enough to the precision of the best atomic
clocks $(\thicksim 2\ast 10^{-17}s)$ and the total number of measurements is
sufficiently close to $\frac{1s}{\Delta T}$ , then the deviations from
formula (10) will occur. \ It is of paramount importance that the number of
measurements must be large, close enough to $\frac{1s}{\Delta T}$ . \ An
instrument \ performing such a measurement is qualitatively different from
two instruments performing successive measurements with large $\Delta T$. \
This experiment should be performed for the different angles $0<\theta <\pi $
changing rapidly.

If the probability for the direction $z$ of spin is $1$ (device A), then for
the direction at the angle $\theta $ the probability is $\mid \cos \frac{1}{2%
}\theta \mid ^{2}$. \ Our hypothesis asserts that `eye for the component of
the spin' apparatus described schematically above will yield a different
curve. \ At least, experiment should be performed for $\theta =\frac{1}{2}%
\pi ,\pi $ because the respective probabilities $\frac{1}{2},0$ are best
suited for statistical tests.

We put forward this hypothesis motivated by the genuine fortuitousness
principle. \ It is fruitless to speculate on a character of deviations
without experimental data. \ Probably, they are `irregular' and very
`quantum'.

However, a few words are needed about the choice of quantum system. \ We
recommend a proton because it is well known stable particle, and its mass $%
(938.27MeV)$ is much greater than that of an electron $(0.51MeV)$ . \ We
consider `point-like' electron and relativistic photon much more elusive. \
Our instruments are governed by laws of electromagnetism. \ One can
speculate that critical time $\Delta T$ should surpass $\frac{\text{Bohr
radius}}{c}\cong 2\ast 10^{-19}$ by factor $10^{2}$ , which accounts for
complexity of an instrument.

Our proposal does not assume the existence of hidden variables, new type of
interaction or new particles. \ We expect a breakdown of QM because
probability measure (2) is not fine enough for the proposed instrument. \ On
the other hand, qualitative features of QM are preserved.

We finish with the remark that our hypothesis has nothing to do with the
putative relaxation processes of proton (proton as an entity) happening
after the measurement. \ Acausal clicks are beyond such processes.

\bigskip

\textbf{References}

1. Araki, Huzihiro (1999), \textit{The Mathematical Theory of Quantum
Fields, }Oxford University Press

2. Bohr, A., Mottelson, B. R., and Ulfbeck, O. ,(2004) \textquotedblleft The
Principle Underlying Quantum Mechanics\textquotedblright , \textit{%
Foundations of Physics, }Vol. 34, No. 3, 405 - 417

3. Bub, Jeffrey (1988), \textquotedblleft How to Solve the Measurement
Problem of Quantum Mechanics\textquotedblright , \textit{Foundations of
Physics}, Vol. 18, No. 7, 701 - 722

4. Caves, C. , and Schack, R. (2004), \textquotedblleft Properties of the
frequency operator do not imply the quantum probability
postulate\textquotedblright \textit{, }arXiv:quant-ph/0409144

5. Mazzucchi, S. (2000), \textquotedblleft On the Observables Describing a
Quantum Reference Frame\textquotedblright , arXiv:quant-ph/0006060

6. Sewell, G. (2002), \textit{Quantum Mechanics and Its Emergent Macrophysics%
}, Princeton University Press

7. Sewell, G. (2005), \textquotedblleft On Mathematical Structure of Quantum
Measurement Theory\textquotedblright , arXiv:math-ph/0505032

8. Toller, M. (1998), \textquotedblleft Localization of Events in
Space-Time\textquotedblright , arXiv:quant-ph/9805030

9. Wigner, E. P. (1939), \textquotedblleft On Unitary Representations of the
Inhomogeneous Lorentz Group\textquotedblright , Ann. Math. \textbf{40}, 149
- 204

10. Wigner, E. P. (1959), \textit{Group Theory and Its Applications to the
Quantum Mechanics of Atomic Spectra}, Academic Press

\end{document}